\documentclass[a4paper]{article}
\usepackage{url}
\usepackage[utf8]{inputenc}
\usepackage[english]{babel}
\usepackage{graphicx}
\usepackage{mathtools}
\usepackage{booktabs}
\usepackage{multirow}
\usepackage{cite}
\usepackage[left=1.25in, right=1.25in, top=1.25in, bottom=1.25in]{geometry}

\graphicspath{ {./images/} }

\author{
	Jonathan Sadighian\\
	 \textit{\'Ecole Pour l'Informatique et les Techniques Avanc\'ees (EPITA)}
	\\
	\texttt{jonathan.m.sadighian@gmail.com}
}

\title{Deep Reinforcement Learning in Cryptocurrency Market Making}

\begin{document}
	\maketitle
	
	\begin{abstract}
		This paper sets forth a framework for deep reinforcement learning as applied to market making (DRLMM) for cryptocurrencies. Two advanced policy gradient-based algorithms were selected as agents to interact with an environment that represents the observation space through limit order book data, and order flow arrival statistics. Within the experiment, a forward-feed neural network is used as the function approximator and two reward functions are compared. The performance of each combination of agent and reward function is evaluated by daily and average trade returns. Using this DRLMM framework, this paper demonstrates the effectiveness of deep reinforcement learning in solving stochastic inventory control challenges market makers face.
	\end{abstract}


	\section{Introduction} \label{Introduction}
	In the past 10 years, algorithmic trading has been gaining market share and currently represents approximately 70 percent of trading activity for equity markets in the United States 
	\cite{CoherentMarketInsights}. Algorithmic trading takes place in quote-driven electronic marketplaces, where participants purchase or sell securities by submitting orders to an exchange. Exchanges maintain orders in a centralized limit order book (LOB), which arranges orders by direction (i.e., sell or buy), price, and time-queue priority. Exchange market makers facilitate transactions amongst participants by providing liquidity in the LOB. Traditionally, statistical approaches have been used to model limit order books for market making and inventory control optimization \cite{bacry2013hawkes, Law2019MarketMU}. Understanding that there have been many recent advancements in deep reinforcement learning in other domains such as video games \cite{mnih2015humanlevel, OpenAI_dota}, yet very little research published on the application of deep reinforcement learning to market making, the focus of this paper is to apply deep reinforcement to cryptocurrency market making.
	
	This paper is structured as follows: section \ref{Introduction}, preliminaries on market making, limit order books, and reinforcement learning; section \ref{RelatedWork}, related research; section \ref{Contributions}, contributions in this paper; section \ref{ExperimentDesign}, experiment design; section \ref{Methodology}, experiment methodology; section \ref{Results}, experiment results and analysis; and section \ref{Conclusion}, conclusion and future work.

	\subsection{Market Making (MM)}
	A market maker’s purpose is to provide liquidity to a marketplace and facilitate transactions amongst participants. This requires continuously maintaining two things: bid and offer orders at a range of prices, and an inventory of positions. Market makers generate revenue by capturing the spread (i.e., the difference between the best bid and ask price). To be profitable, market makers must continuously update price quotes to reflect changing market conditions, and manage their inventory of positions to avoid overexposure to a directional move in the market. This stochastic optimization problem of maintaining quotes and inventory levels poses an interesting challenge for reinforcement learning.
	
	Market makers operate differently based on asset class. Equity exchanges in the United States, such as the NYSE or AMEX, have a designated market maker for each security, who in turn has an obligation to provide a specified amount of liquidity to that security. Other exchanges in the United States, such as NASDAQ, have more than one designated market maker per security. However, in cryptocurrency exchanges, such as Coinbase or Bitfinex, there are no designated market makers. In these marketplaces, liquidity is provided by participants, who can be either institutional or retail investors.

	\subsection{Limit Order Books (LOB)}
	When participants in cryptocurrency markets want to buy or sell an asset, they must send their order to an exchange. If the participant wants to buy (or sell) an asset at the current market price, they can submit a market order, whereas if the participant has a specific price limit they are willing to pay for the asset, they can send a limit order. Exchanges use a limit order book to track inventory of all limit orders accepted by the exchange. 
	
	LOBs are grouped by order direction (i.e.,  buy or sell), price, and time priority (i.e., FIFO order). The bid side of the LOB contains all the buy orders, whereas the ask side contains all the sell orders. If an order within the LOB is matched with a market order, it is removed from the LOB upon fulfillment. In this scenario, the participant who made the limit order is referred to as the liquidity provider or maker, and the participant who sent the market order is the liquidity remover or taker. Since exchanges profit from trading volume, most electronic exchanges use a maker-taker transaction fee model. Under this fee structure, makers are incentivized to provide liquidity to a given exchange in the form of a rebate, or reduced transaction fee. Market makers usually rely on rebates as part of their trading strategies to prevent transaction fees from eroding profitability.

	\subsection{Order Flow Imbalances (OFI)}
	The number and frequency of orders received by the LOB can provide additional information relating to future prices of an asset \cite{wu2019queuereactive, Xu2019MultiLevelOI}. In this paper, orders are classified by the following three categories as described in Table \ref{table:ofi}: market orders, limit orders, and cancellation orders. Under this schema, market and cancel orders remove liquidity from the LOB, while limit orders add liquidity. Throughout the rest of this paper, the order flow statistics for these events are referred to as order flow imbalances. 
	
	\begin{table}[h	t]
		\centering
		\begin{tabular}{p{0.02\textwidth} p{0.30\textwidth} p{0.6\textwidth}}
			\toprule\toprule
			\textbf{\#} & \textbf{Order Event Type} & \textbf{Description} \\ [0.5ex] 
			\midrule
			1 & Cancel, $C$ & Order cancellation request $C$ received by the exchange, resulting in liquidity being removed from the LOB \\ 
			\midrule
			2 & Limit, $L$ & New limit order $L$ received by the exchange, resulting in liquidity being added to the LOB \\
			\midrule
			3 & Market, $M$ & Market order $M$ received by the exchange, resulting in liquidity being removed from the LOB when matched with a resting limit order \\ [1ex]
			\midrule
		\end{tabular}
		\caption{Types of order flow arrival events}
		\label{table:ofi} 
	\end{table}

	\subsection{Trade Flow Imbalance (TFI)}
	Trade flow imbalances quantify the ebbs-and-flows of transactions for a given asset. It has been demonstrated by \cite{Cont_2013} that trade flow data can be used as an indicator to predict price movements. Typically, the TFI is calculated via the net difference between notional values of buy and sell transactions for a given time period. In this paper, the TFI is calculated differently, as described in equation \ref{eq:tfi} in section \ref{ess}.

	\subsection{Reinforcement Learning (RL)}
	Reinforcement learning is a branch of machine learning wherein learning occurs through trial and error, rather than by example, as in supervised learning. As such, RL algorithms are able to learn goals, rather than patterns. RL problems are modeled using the Markov Decision Process (MDP), a framework that enables agents to learn sensations, actions, and goals. Learning is achieved by maximizing a reward signal through trial and error \cite{Sutton1998}.
	
\begin{figure}[h!]
		\hfill \includegraphics[scale=0.2]{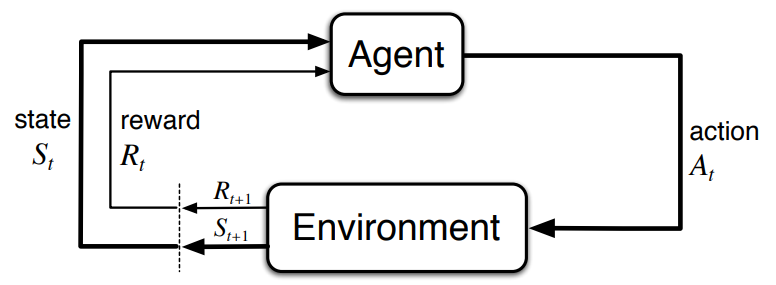} \hspace*{\fill}
		\caption{Markov Decision Process represented as a loop}
		\label{fig:mdp}
	\end{figure}

	The MDP framework is expressed by the tuple $\langle S, A, R, P \rangle$ and is depicted in figure \ref{fig:mdp}:
	\begin{itemize}
		\item State space, $S$, is the set of all valid states (what the agent can see),
		\item Action space, $A$, is the set of all valid actions (what the agent can do in response to an observation),
		\item Reward, $R$, is the reward (feedback signal from the environment), and
		\item Policy, $P$, is the map of actions to a given state.
	\end{itemize}

	In context of market making, the MDP framework is referred to as a Partially Observable Market Decision Process (POMDP), since the agent cannot have a complete view of the state space. In a POMDP, the partially observable state space is referred to as an observation.
	
	In model-free reinforcement learning, the agent has no prior knowledge of its environment and learns to solve problems through trial-and-error, in which the agent iteratively updates its policy while collecting experience from environmental interactions. There are various ways in which model-free algorithms can update their policies: value-based methods optimize the policy indirectly through estimating action-values; policy-gradient methods optimize the policy directly without requiring a value function; methods that learn approximations to both policy and value functions are often called actor–critic methods, where “actor” is a reference to the learned policy, and “critic” refers to the learned value function (usually a state-value function) \cite{Sutton1998}. The algorithms used in this experiment are model-free actor-critic algorithms.

	\subsection{Deep Reinforcement Learning (DRL)}
	In Deep Reinforcement Learning (DRL), the agent’s policy  is an artificial neural network, into which the environment observation is input,  triggering an output of the action the agent should take. DRL enables the agent to generalize more effectively than other function approximators for high-dimensional spaces, such as linear approximation \cite{Sutton1998}.

	\section{Related Work} \label{RelatedWork}
	There has been substantial recent research concerning the successful application of deep reinforcement learning algorithms to complex tasks, such as the game of Go \cite{SilverHuangEtAl16nature} or Atari \cite{mnih2015humanlevel, Espeholt2018IMPALASD}. The aforementioned experiments have implemented value-based algorithms to optimize the agent’s policy indirectly through an action-value function, whereas this paper focuses on policy-gradient methods.
	
	In financial services, it has been demonstrated that a model-free approach with policy gradient methods can be effective at trading or portfolio management and that a policy gradient method is more effective at learning the dynamics of trading than Q-learning, a value-based method \cite{MoodyRLT1998, MoodyRLT1999, MoodyRLT2001, Gold2003FXTV, lu2017agent}. Other findings show that using risk-based metrics, such as the Sharpe Ratio (or other variations like Differential Sharpe Ratio), for the reward function (i.e., feedback signal) outperform algorithms that use the more intuitive profit-and-loss reward function \cite{MoodyRLT2001, Gold2003FXTV, lu2017agent}. However, none of these papers use data from limit order books or trade and order flow data as an input into the agent’s observation space, but rather price returns.
	
	Within the research community, there is a dearth of published work exploring the application of DRL to automated market making. \cite{spooner2018market} created a framework for market making and evaluated several value-based reinforcement learning algorithms to achieve a stable performance \emph{without} using a neural network. They used the top five levels of the LOB in combination with a few other hand-crafted features to represent the agent’s observation space and a linear combination of tile codings as a function approximator. They attribute success to their custom reward function and linear combination of tile codes, with the help of eligibility traces. \cite{patel2018optimizing} proposed a macro-micro duel agent architecture for solving the market making problem, wherein the macro-agent views higher-resolution data and generates trade signals, and the micro-agent views the LOB and is responsible for order execution. Although \cite{patel2018optimizing}’s agent outperformed their baseline, the lack of inventory constraints on the agent makes the results impractical for live trading. Unlike these approaches, this paper uses more levels from the LOB, trade and order flow imbalances, and advanced policy-gradient algorithms to achieve stable performance for cryptocurrency market making.
	
	There has been ample research in modeling the LOB using statistical, machine learning, and deep learning approaches. \cite{bacry2013hawkes, Law2019MarketMU, wu2019queuereactive, morariupatrichi2018statedependent} use a marked point process to indirectly predict midpoint price changes and inventory control for market making. \cite{Xu_2019} use order flow imbalances and linear regressions to predict midpoint changes. \cite{Zhang_2019} use a custom stack of convolutional and LSTM neural networks with an inception module to predict midpoint price changes. Unlike these stochastic and supervised machine learning approaches, which rely on heuristics to create a trading system, this paper uses reinforcement learning with neural networks to learn an end-to-end automated trading solution for market making.

	\section{Contributions} \label{Contributions}
	The main contributions of this paper are the analyses of the DRLMM frameworking using advanced policy gradient-based algorithms. In contrast to existing research \cite{spooner2018market}, this experiment features an expanded scope, including 15 LOB levels, in addition to trade and order flow arrival data for the agent’s exploitation. This paper aims to provide the first published research using LOB, TFI and OFI to create an informed deep reinforcement learning market making agent.	
	
	The main contributions are detailed below:
	\begin{enumerate}
		\item \textbf{Analysis of advanced policy gradient-based algorithms:}
		The performance and generalization capability across three cryptocurrency pairs are compared for Advantage Actor Critic and Proximal Policy Optimization algorithms.
		\item \textbf{Reward functions for market making:}
		Two reward functions, positional profit-and-loss and trade completion, are evaluated in context of deep reinforcement learning applied to market making.
		\item \textbf{Framework for a DRLMM agent:}
		Agents that are able to effectively learn the end-to-end process of market making with no prior knowledge are demonstrated.
	\end{enumerate}

	\section{Experiment Design} \label{ExperimentDesign}

	\subsection{Environment Design}
	The agent’s observation space is a combination of three sub-spaces: the environment state space (ESS), consisting of LOB, TFI and OFI snapshots with a lookback window size $w$; the agent state space (ASS), consisting of handcrafted risk and position indicators; and the agent action space (AAS), consisting of a one-hot vector of the agent’s latest action.

	\subsubsection{Environment State Space (ESS)} \label{ess}
	In this paper, the ESS consists of eight indicators: 1) LOB levels rendered as stationary prices; 2) cumulative notional value at each price level; 3) imbalances for cumulative notionals; 4) order inflow imbalances at LOB levels; 5) the spread between the best bid and ask prices; 6) custom price momentum indicator; 7) custom trade flow imbalance indicator; and 8) environmental reward signal.
	
	\paragraph{Price Level Distance to Midpoint} \label{PriceLevelDistance}
	Distance $\xi$ is the difference between price $p$ at LOB level $i$ from midpoint $m$ at time $t$. The LOB’s first 15 price levels are extracted from $bid$ and $ask$ sides of the LOB, thus creating 30 values for this feature.
	
	\begin{equation}
		\xi^{bid, ask}_{t, i} = \frac{p^{bid, ask}_{t, i}}{m_{t}} - 1
	\end{equation}

	\paragraph{Cumulative Notional Value at Price Level}
	Cumulative sum $\chi$ is the sum of the product of price $p$ and quantity $q$ at LOB level $i$ at time $t$ for both $bid$ and $ask$ sides. The LOB’s first 15 price levels are extracted from $bid$ and $ask$ sides of the LOB, thus creating 30 values for this feature.
	
	\begin{equation}
		\chi^{bid, ask}_{t, i} =\sum_{i=0}^{I-1} p^{t, i}_{bid, ask} \times q^{t, i}_{bid, ask}
	\end{equation}
	
	\paragraph{Notional Imbalances}
	The order imbalances $\iota$ are represented by notional value and normalized to the scale $[-1, 1]$. Since there are 15 price levels included, there are 15 values for this feature.
	
	\begin{equation}
		\iota_{t, i} = \frac{\chi^{ask, q}_{t, i} - \chi^{bid, q}_{t, i}}{\chi^{ask, q}_{t, i} + \chi^{bid, q}_{t, i}}
	\end{equation}
	
	\paragraph{Order Flow Imbalances (OFI)} \label{OrderFlowImabalances}
	The sum of notional values for cancel $C$, limit $L$, and market $M$ orders is captured between each LOB snapshot. Since $i$ is 15 in this experiment, this feature is represented by a vector of 30 values, 15 per each side of the LOB.
	
	\begin{equation}
		OFI_{t, i}^{bid, ask} = OFI(L)_{t, i}^{bid, ask} - OFI(C)_{t, i}^{bid, ask} - OFI(M)_{t, i}^{bid, ask}
	\end{equation}
	
	where
	$OFI(C)^{bid, ask}_{t, i} = C^{bid, ask, p}_{t, i} \times C^{bid, ask, q}_{t, i}$
	and
	$OFI(L)^{bid, ask}_{t, i} = L^{bid, ask, p}_{t, i} \times L^{bid, ask, q}_{t, i}$
	and
	$OFI(M)^{bid, ask}_{t, i} = M^{bid, ask, p}_{t, i} \times M^{bid, ask, q}_{t, i}$.
	
	\paragraph{Trade Flow Imbalances (TFI)}
	The Trade Flow Imbalances $TFI$ indicator measures the magnitude of buyer initiated $BI$ and and seller initiated $SI$ transactions over a given window $w$. This indicator is scaled to the range of $[-1, 1]$ and is applied to 3 different windows (5, 15, and 30 minutes), thus is represented by a vector of 6.
	
	\begin{equation} \label{eq:tfi}
		TFI_{t} = \frac{gain_{t} - loss_{t}}{gain_{t} + loss_{t}}
	\end{equation}
	
	where:
	$gain_{t} = \sum_{n=0}^{w} BI_{n}$
	and
	$loss_{t} = \sum_{n=0}^{w} SI_{n}$.
	
	\paragraph{Spread}
	The spread  is the difference between the best bid $p^{bid}$ and best ask $p^{ask}$.	

	\begin{equation}
		\varsigma_{t} = p_{t}^{bid} - p_{t}^{ask}
	\end{equation}
	
	\paragraph{Custom RSI}
	The relative strength index indicator (RSI) measures the magnitude of prices changes over a given window $w$. This custom implementation $CRSI$ replaces the default $mean$ calculation with a $sum$, and scales the output to the range $[-1, 1]$. This indicator is applied to three different windows (5, 15, and 30 minutes), and is thus represented by a vector of 3 features.
	\begin{equation}
	CRSI_{t} = \frac{gain_{t} - \arrowvert loss_{t} \arrowvert}{gain_{t} + \arrowvert loss_{t} \arrowvert}
	\end{equation}	
	where	
	$gain_{t} = \sum_{n=0}^{w} \Delta m_{n} \textnormal{ if } \Delta m_{n} > 0 \textnormal{ else } 0$
	and	
	$loss_{t} = \sum_{n=0}^{w} \Delta m_{n} \textnormal{ if } \Delta m_{n} < 0 \textnormal{ else } 0$
	and	
	$\Delta m_{t} = \frac{m_{t}}{m_{t-1}} - 1$.
	
	\paragraph{Reward}
	The reward $r$ from the environment, as described in section \ref{RewardFunction}.

	\subsubsection{Agent State Space (ASS)}
	The ASS provides transparency into the agent’s operations. In this paper, the ASS consists of five indicators: 1) percentage of inventory utilized; 2) total profit-and-loss; 3) unrealized profit-and-loss; 4) distance for each open order to midpoint price; 5) a custom function to measure the progress of an order’s position in the queue and partial executions.
	
	\paragraph{Inventory Ratio}
	The agent’s inventory ratio $\upsilon$ is the inventory count $ic$ represented as a percentage of the maximum inventory $IM$, scaled between $[0, 1]$, and represented as a vector with 2 values.	
	\begin{equation}
		\upsilon_{t}^{long, short} = \frac{ic^{long, short}}{IM}
	\end{equation}
	
	\paragraph{Total PnL}
	The agent’s total profit-and-loss $TPNL$ is the sum of realized and unrealized profit and losses. In this experiment, the $TPNL$ is scaled by a scalar value $\rho$, which represents the daily PnL target. The current midpoint $m$ at time $t$ marks the unrealized position to market value. Total PnL is represented as a scalar.
	\begin{equation}
		TPNL_{t} = \frac{PNL_{t}^{realized} + PNL_{t}^{unrealized}}{\rho}
	\end{equation}
	
	\paragraph{Unrealized PnL}
	The agent’s current unrealized PnL $PNL_{unrealized}$ is the sum of unrealized PnL across all open positions. The unrealized PnL feature is represented as a vector with 2 values, containing both $\textnormal{long}$ and $\textnormal{short}$ positions.
	\begin{equation}
		PNL_{unrealized}^{long, short} = \frac{p_{average}^{long, short}}{m} - 1
	\end{equation}
	where $m$ is the midpoint price at time $t$.
	
	\paragraph{Limit Order Distance to Midpoint}
	The agent’s limit order distance to midpoint is the distance $\zeta$ of the agent’s open $bid$ and $ask$ limit orders $L$ to the midpoint price $m$ at time $t$. The feature is represented as a vector with 2 values.
	\begin{equation}
		\zeta^{long, short}_{t} = \frac{L^{bid, ask}_{t}}{m} - 1
	\end{equation}
	
	\paragraph{Order Completion Ratio}
	Order completion $\eta$ is a custom indicator that incorporates an order’s relative position in the LOB queue $q$ and partial executions $ex$ relative to the order size $sz$. The feature scales the value to the range $[-1, 1]$ and is represented as a vector with 2 values.	
	\begin{equation}
		\eta_{t}^{long, short} = \frac{ex_{t}^{long, short} - q_{t}^{long, short}}{q_{t}^{long, short} + sz_{t}^{long, short}}
	\end{equation}

	\subsubsection{Agent Action Space (AAS)}
	The agent action space $A$ consists of 17 possible actions, as outlined in Table \ref{table:action_space}. The idea is that the agent can take four general actions: do nothing, place its orders asymmetrically in the LOB, skew its orders to either $buy$ or $sell$ sides of the LOB, or flatten its inventory of positions.
	
	\begin{table}
		\centering
		\label{table:action_space}		
		\begin{tabular}{lccccccccccccccc}
			\toprule\toprule
			Action ID & 2 & 3 & 4 & 5 & 6 & 7 & 8 & 9 & 10 & 11 & 12 & 13 & 14 & 15 & 16 \\
			\midrule
			Bid & 0 & 0 & 0 & 4 & 4 & 4 & 4 & 9 & 9 & 9 & 9 & 14 & 14 & 14 & 14 \\
			Ask & 4 & 9 & 14 & 0 & 4 & 9 & 14 & 0 & 4 & 9 & 14 & 0 & 4 & 9 & 14 \\
			\midrule
			Action 1 & \multicolumn{15}{l}{No action} \\
			Action 17 & \multicolumn{15}{l}{Market order $M$ with size $ic$} \\
			\bottomrule\bottomrule
		\end{tabular}
		\caption{The agent action space with 17 possible actions. The numbers in the $Bid$ and $Ask$ rows represent the price level at which the agent's orders are set to for a given action.}
	\end{table}

	\subsubsection{Reward Function} \label{RewardFunction}
	There are two reward functions evaluated in this paper as part of the DRLMM framework: positional PnL and trade completion. Each function returns a reward $r$, which is the agent's feedback signal the agent uses to learn.
	
	\paragraph{Positional PnL} \label{PositionalPnL}
	The reward $r$ is the change in the agent’s unrealized position value, plus any realized gains or losses from the current step $t$. If no positions have been closed between time steps $t$ and $t-1$, the realized gains input is set to 0.
	
	\begin{equation}
		r_{t} = \Delta m_{t} \times ic_{t} + PNL_{t}^{realized}
	\end{equation}	
	
	where the midpoint price return is $\Delta m_{t} = \frac{m_{t}}{m_{t-1}} - 1$.
	and inventory count $ic$ is the number of positions held in inventory at time $t$.
	
	\paragraph{Trade Completion} \label{TradeCompletion}
	The reward $r$ is a value between $[-1, 1]$ that is derived from a clipped goal-based custom reward function. This reward function only generates a value other than 0 when a position is closed and realized PnL is generated. If the realized PnL is greater (or less) than a predefined threshold, the reward $r$ is 1 (or -1) otherwise, if the realized PnL is in between the thresholds, the realized PnL in percentage terms is the reward. Using this clipped approach, the agent is encouraged to open and close positions with a targeted profit-to-loss ratio, and is not rewarded for longer term price speculation.
	
	\begin{equation}
	    r_{t} = 
		\begin{cases}
			1, & \textnormal{if } PNL^{realized} \geq \epsilon * \varpi \\
			-1, & \textnormal{if } PNL^{realized} \leq - \varpi \\
			PNL^{realized}, & \textnormal{otherwise} \\
		\end{cases}
	\end{equation}
		
	where $\epsilon$ is a constant used for the multiplier and $\varpi$ is a constant used for the threshold.
	
	\begin{figure}[h!]
		\hfill \includegraphics[scale=0.50]{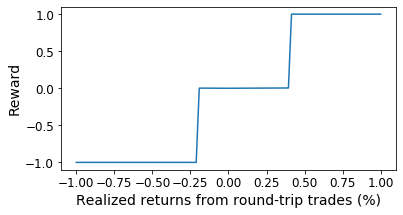} \hspace*{\fill}
		\caption{Trade Completion reward function for returns ranging from -1\% to 1\% when $\epsilon$ is set to 2 and $\varpi$ is set to 0.2\%.}
		\label{fig:trade_completion_reward_function}
	\end{figure}

	\subsection{Function Approximator}
	The function approximator is a multilayer perceptron (MLP), which is a forward feed artificial neural network. The architecture in this implementation consists of 3-layer network with a single shared layer for feature extraction, followed by separate policy and value networks.

	\begin{figure}[h!]
		\hfill \includegraphics[scale=0.1]{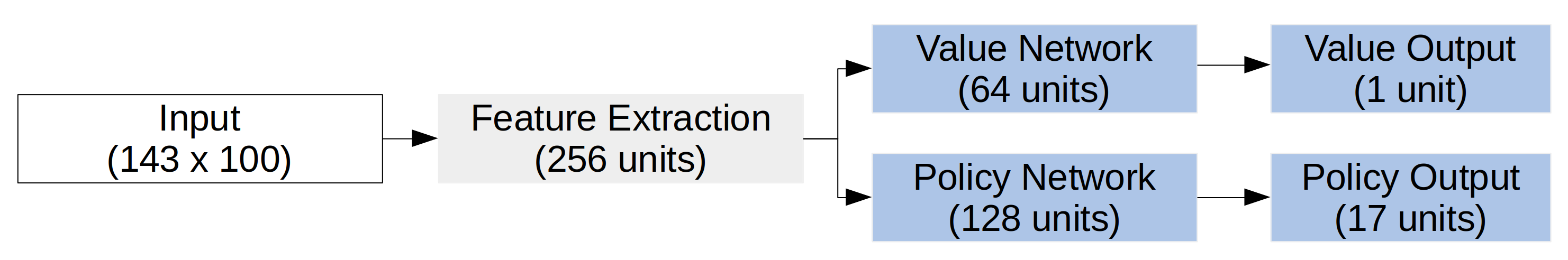} \hspace*{\fill}
		\caption{Architecture of MLP neural network used in experiment. The input space is a matrix of 143 features with 100 lags. \emph{Gray} represents shared layers. \emph{Blue} represents non-shared layers.}
		\label{fig:neural_network}
	\end{figure}

	\subsection{Agents}
	Two advanced policy gradient methods are used as market making agents: Advantage Actor-Critic (A2C) and Proximal Policy Optimization (PPO). Since both learning algorithms run on multiple processes, they require nearly the same amount of time to train. The same policy network architectures are used in each experiment, and hyperparameters are listed in the appendix (section \ref{AgentConfigurations}). It is worth noting that Actor-Critic with Experience Replay (ACER) was originally in scope for this paper, but was de-scoped due to poor performance after demonstrating instability and slow learning in comparison to A2C and PPO.
	
	\subsubsection{Advantage Actor-Critic (A2C)}
	The A2C is an on-policy model-free actor-critic algorithm that is a descendent of policy gradient methods. It interacts with the environment asynchronously while maintaining a policy $\pi (a_{t}|s_{t}; \theta)$ and estimate of the value function $V(s_{t}; \theta_{v})$, and synchronously updates parameters to improve GPU efficiency compared to sibling algorithm A3C \cite{mnih2016asynchronous}. Being a more evolved algorithm than a vanilla policy gradient, the A2C algorithm uses its value function to reduce variance and improve stability when learning through calculating the advantage $A(s_{t} | a_{t})$ at each state. The advantage is the difference between the action-value $Q(s_{t} | a_{t})$ and value function $V(s_{t})$ (i.e., the advantage of taking a particular action at a given state). The A2C algorithm also uses n-step returns to update both policy and value-function, which results in more stable learning than a vanilla policy gradient, which uses 1-step returns.
	
	The A2C update is calculated as	
	\begin{equation}
		\nabla_{\theta} J(\theta) = \nabla_{\theta ‘} log \pi (a_{t} | s_{t}; \theta ') A(s_{t}, a_{t}; \theta, \theta_{\upsilon})	
	\end{equation}
	
	where $A(s_{t}, a_{t}; \theta, \theta_{\upsilon})$ is the estimate of the advantage function given by $\sum_{i=0}^{k-1} = \gamma^{i} r_{t + i}  + \gamma^{k} V(s_{t + k}, \theta_{\upsilon}) - V(s_{t}, \theta_{\upsilon})$, where $k$ can vary by the upper bound $t_{max}$ \cite{mnih2016asynchronous}.
	
	\subsubsection{Proximal Policy Optimization (PPO)}
	The PPO is an on-policy model-free actor-critic algorithm and a descendent of policy gradient methods. It interacts with the environment asynchronously while indirectly optimizing a policy $\pi_{\theta}(a_{t} | s_{t})$ through a clipped surrogate function $L^{CLIP}$ \cite{schulman2017proximal} that represents the difference between the new policy after the most recent update $\pi_{\theta}(a | s)$ and the old policy before the most recent update $\pi_{\theta_{k}}(a | s)$. Although this algorithm descends from policy gradient methods, it is different in that the surrogate function, not the policy, is optimized directly. This surrogate function removes the incentive for a new policy to depart from the old policy, thereby increasing learning stability. Like the A2C, the PPO algorithm updates its parameters $\theta$ synchronously.
	
	The PPO Clip update is calculated as		
	\begin{equation}
		\small L(s,a,\theta_k,\theta) = \min\left( \frac{\pi_{\theta}(a_{t}|s_{t})}{\pi_{\theta_k}(a_{t}|s_{t})} A^{\pi_{\theta_k}}(s_{t},a_{t}), \;\;
		\textnormal{clip}\left(\frac{\pi_{\theta}(a_{t}|s_{t})}{\pi_{\theta_k}(a_{t}|s_{t})}, 1 - \epsilon, 1+\epsilon \right) A^{\pi_{\theta_k}}(s_{t},a_{t}) \right)
	\end{equation}
	
	\normalsize where $\epsilon$ is a hyperparameter constant \cite{schulman2017proximal}.

	\section{Methodology} \label{Methodology}
	
	\subsection{Data Collection}
	LOB data for cryptocurrencies is free to access via WebSocket, but not readily downloadable from exchanges, and therefore requires recording. The data set for this experiment was recorded using level-3 tick data from the Coinbase exchange, persisted into a Mongo database, and replayed to fully reconstruct the LOB and derive TFI and OFI metrics. During data replay, snapshots of the LOB, TFI and OFI statistics were taken every n-intervals and exported to a CSV file to reduce the computational burden of parsing millions of ticks per trading day. In this experiment, the snapshots were taken every 1-seconds, resulting in approximately 84,600 records per trading day. 
	
	\subsection{Data Normalization}
	Since LOB data cannot be used for machine learning without preprocessing, it is necessary to apply a normalization technique to the raw data set. For this experiment, the data set was normalized using the approach described by \cite{tsantekidis2018}, which transforms the LOB from non-stationary into a stationary feature set, then uses the previous trading day to fit and $z$-score normalize the current trading day’s values, and in which data point $x$ is $\sigma$ standard deviations from the mean $\bar{x}$.
	
	\begin{equation}
	z_{x} = \frac{x - \bar{x}}{\sigma}
	\end{equation}
	
	\subsection{Environment Rules}
	The environment invokes business rules to make the simulation as realistic as possible. For example, agents are permitted to hold only one open order at a time on each side of the LOB (the ask and bid sides). Once an order is filled, the agent must wait until the next environment step to select an action (such as replenishing the filled order in the order book). Additionally, agents are able to update partially filled open orders with new price levels, and upon order completion, the average execution price is updated in all PnL or risk calculations. If the agent has an open position on one side of the book and fills an open order on the opposing side, the existing position is netted in FIFO for PnL calculations. 
	
	Each time a new order is posted to the LOB, its price and standing in the queue is updated according to relative notional value (e.g. price, quantity). If the agent posts a new order at the same price as existing resting orders, the new order price jumps a one-tick increment ahead of the queue by default, assuming there is no liquidity posted at that price level. However, it is important to note that this hard-coded order posting strategy is less relevant for live trading, wherein other market participants have the ability to repost their own orders ahead of the agent’s, therby triggering a series of requotes.
	
	In general, transaction fees for limit orders are not taken into account. Since limit order fees or rebates vary by exchange, transaction fees are only applied to market orders (0.20\% per side), which the agent can engage by selecting the “flatten inventory” action.
	
	\subsection{Training and Testing}
	The market-making agents (A2C and PPO) were trained on 8 days of data (September 27 to October 4, 2019) and tested on 2 trading days (November 1 and 2, 2019) using three cryptocurrency pairs: Bitcoin (BTC-USD), Ether (ETH-USD), and Litecoin (LTC-USD). Each trading day consists of 84,600 snapshots, which reflects 1 snapshot per second in a 24-hour market. Each agent’s performance is evaluated on both total daily and average trade returns. 
	
	In each experiment, agents are trained for 10 million steps using 5 action repeats, enabling agents to accelerate learning and take 50 million environmental steps. It is important to note that during action repeats, the agent’s action is only performed once on the first step, with all subsequent steps consisting of “no action,” thereby avoid performing illogical repetitive actions multiple times in a row, such as “flatten all inventory.” Each agent uses the hyperparameters specified in the appendix \ref{AgentConfigurations}.

	\section{Results} \label{Results}
	In this section, the performance of each agent (PPO and A2C) are compared on the testing data set. The performance of each data set is evaluated in terms of daily returns (Table \ref{table:DailyReturns}), as well as the average trade return (Table \ref{table:AverageTrade}). Additionally, the agents are tested on different cryptocurrency pairs to evaluate their capability to generalize. Both agents learned how to generate daily profits and manage inventories using the DRLMM framework.

	\begin{table}\label{table:AverageTrade}
		\centering
		\renewcommand{\arraystretch}{1.2}
		\begin{tabular}{|p{0.2\textwidth}|p{0.1\textwidth}|p{0.1\textwidth}|p{0.1\textwidth}|p{0.1\textwidth}|p{0.1\textwidth}|p{0.1\textwidth}|}
			\hline
			\multirow{2}{3cm}{\textbf{Reward function}} & \multicolumn{3}{c|}{\textbf{A2C}} & \multicolumn{3}{c|}{\textbf{PPO}} \\
			\cline{2-7}
			& \textbf{\textit{BTC}} & \textbf{\textit{ETH}} & \textbf{\textit{LTC}} & \textbf{\textit{BTC}} & \textbf{\textit{ETH}} & \textbf{\textit{LTC}}\\
			\hline
			Positional PnL & 0.0034\% & -0.0001\% & -0.0002\% & -0.0274\% & 0.0001\% & 0.0005\%  \\ \hline
			Trade Completion & 0.0017\% & 0.0005\% & 0.0017\% & 0.0001\% & 0.0002\% & 0.0018\%  \\ \hline
		\end{tabular}
		\caption{Average of the average per-trade PnL of the \textit{same currency} for two testing days: November 1 and 2, 2019. }
	\end{table}

	\begin{table}\label{table:DailyReturns}
		\centering
		\renewcommand{\arraystretch}{1.2}
		\begin{tabular}{|p{0.2\textwidth}|p{0.1\textwidth}|p{0.1\textwidth}|p{0.1\textwidth}|p{0.1\textwidth}|p{0.1\textwidth}|p{0.1\textwidth}|}
			\hline
			\multirow{2}{3cm}{\textbf{Reward function}} & \multicolumn{3}{c|}{\textbf{A2C}} & \multicolumn{3}{c|}{\textbf{PPO}} \\
			\cline{2-7}
			& \textbf{\textit{BTC}} & \textbf{\textit{ETH}} & \textbf{\textit{LTC}} & \textbf{\textit{BTC}} & \textbf{\textit{ETH}} & \textbf{\textit{LTC}}\\
			\hline
			Positional PnL & 0.0934\% & -0.0851\% & -0.1836\% & -0.1094\% & 0.1320\% & 0.3462\%  \\ \hline
			Trade Completion & 3.3063\% & 0.0643\% & 1.5078\% & 0.4051\% & 0.0650\% & 1.4223\%  \\ \hline
		\end{tabular}
		\caption{Average daily returns of the \textit{same currency} for two testing days: November 1 and 2, 2019. }
	\end{table}

	\subsection{Same Currency} \label{SameCurrency}
\begin{figure}[h!]
		\includegraphics[scale=0.32]{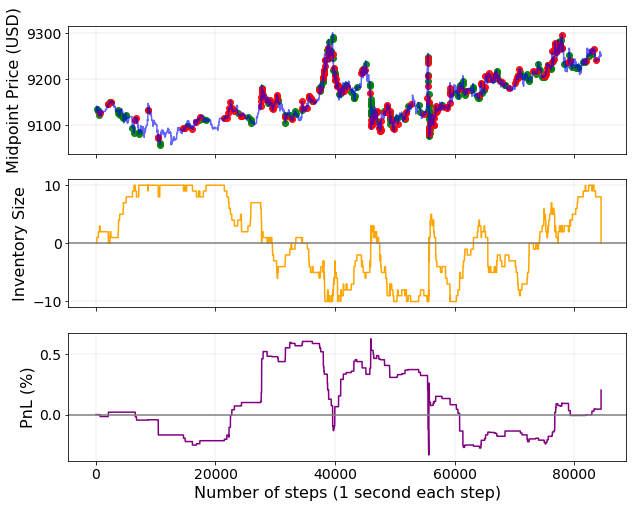}
		\includegraphics[scale=0.32]{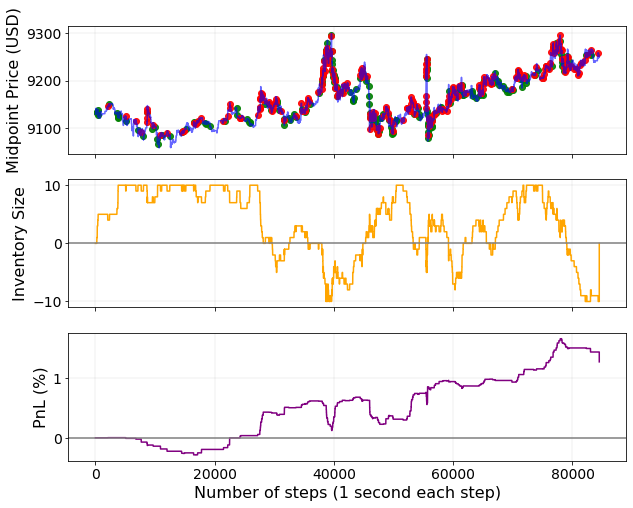}
		\caption{Comparison of action repeat settings for PPO agent trading BTC-USD on November 1, 2019 using trade completion reward function. Buy and sell orders executed by the agent are marked in \emph{red} and \emph{green}. \emph{Left:} Agent with action repeat set to 1. \emph{Right:} Agent with action repeat set to 10.}
		\label{fig:action_repeat_comparison}
	\end{figure}

	The agents achieved positive returns for the evaluation metrics on the testing data sets. BTC and LTC trained agents generated more stable returns than the ETH agent. Interestingly, the performance of each algorithm varies based on the number of action repeats performed during testing, as seen in Figure \ref{fig:action_repeat_comparison}. For example, the A2C algorithm achieves greater returns with a smoother equity curve using smaller action repeat values (i.e, 1 or 5), whereas the PPO algorithm performs better with larger action repeats values (i.e., 5 or 10). In the case of A2C with fewer action repeats, the agent appears to manage its position inventory better, thereby reducing the holding period of each position. For the PPO agent, the opposite effect occurs: The agent accumulates a large position quickly, and holds onto the positions until they become profitable.
	
\begin{figure}[h!]
		\includegraphics[scale=0.32]{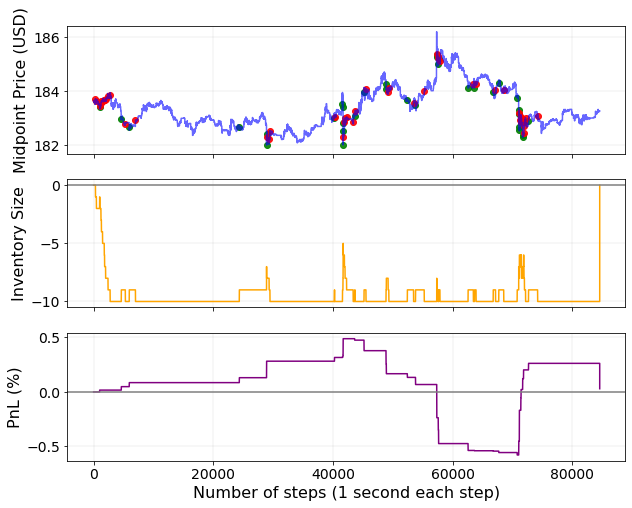}
		\includegraphics[scale=0.32]{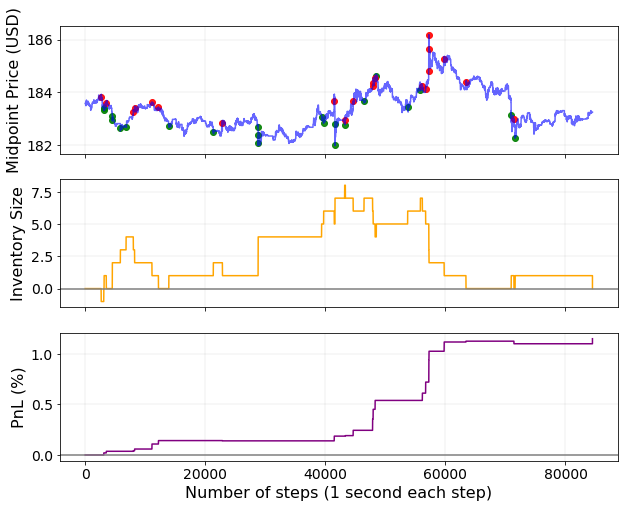}
		\caption{Comparison of reward function effects on trading activities for A2C agent trading ETH-USD on November 2, 2019. Buy and sell orders executed by the agent are marked in \emph{red} and \emph{green}. \emph{Left:} Agent with positional PnL reward function. \emph{Right:} Agent with trade completion reward function.}
		\label{fig:reward_signal_comparison}
	\end{figure}

	As for reward functions, the trade completion approach yielded greater and more stable returns than positional PnL. This is due to trade completion agents generating more trades than positional PnL agents, while also maintaining a lower inventory count, as seen in in Figure \ref{fig:reward_signal_comparison}. Another difference between the two reward functions is the agents' use of the "flatten inventory" action, which instantly liquidates the agent's position. The positional PnL agents used this action more frequently than trade completion agents, which is expected since future changes in midpoint prices directly impact the reward signal.
	
	\subsection{Different Currencies}
\begin{figure}[h!]
		\includegraphics[scale=0.32]{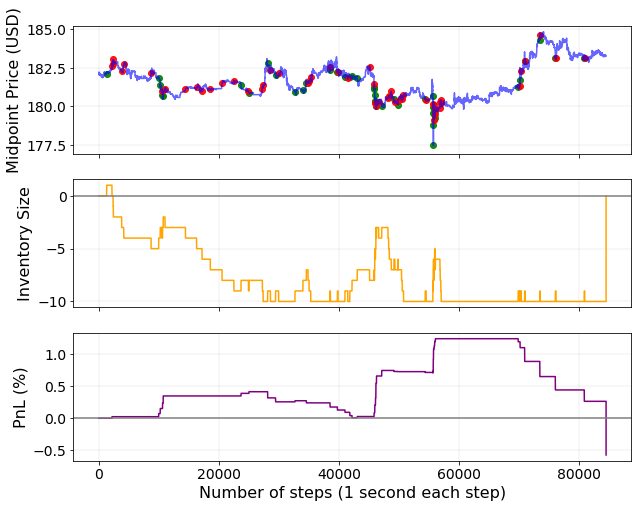}
		\includegraphics[scale=0.32]{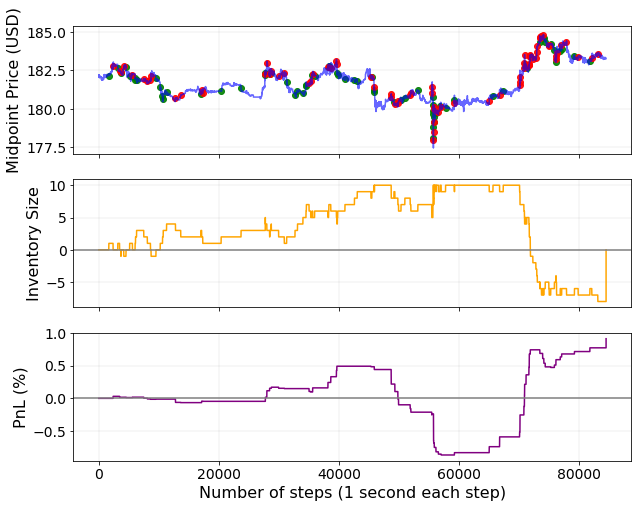}
		\caption{Example of improved performance when PPO agent trades different currency (ETH-USD) from training currency (LTC-USD) on November 1, 2019 using trade completion reward function and action repeat 10. Buy and sell orders executed by the agent are marked in \emph{red} and \emph{green}. \emph{Left:} Agent trained on ETH-USD. \emph{Right:} Agent trained on LTC-USD.}
		\label{fig:different_currency}
	\end{figure}

	When applied to currencies other than the respective training data set, each agent's performance was relatively similar to results in section \ref{SameCurrency}. Some agents trained on different currencies performed better than the \emph{same currency} approach counterparts, as seen in Figure \ref{fig:different_currency}. For example, the PPO agents trained on BTC-USD yielded better performance on ETH-USD data than the agent trained on ETH-USD data. Similar results occurred with the A2C algorithm, where	agent trained on BTC-USD outperformed the agent trained on ETH-USD. These results support the findings of \cite{sirignano2018universal}, where LOBs exhibit universality.
	
	\section{Conclusion} \label{Conclusion}
	This paper applies deep reinforcement learning to cryptocurrency market making using advanced policy gradient methods. The proposed DRLMM framework demonstrates the effectiveness of policy gradient methods in solving stochastic control problems in market making. Using only limit order book data and trade and order flow indicators, the agent is able to generate profitable daily returns without prior knowledge of how the market making process is performed. Moreover, the DRLMM framework enables the agent to generate profitable daily returns on different cryptocurrency data sets, highlighting the agent's ability to generalize effectively and learn the non-linear characteristics of its obervation space.
	
	In future research, there are many other features to explore in the DRLMM framework. It would be beneficial to compare additional reward functions, such as Differential Sharpe Ratio \cite{MoodyRLT1999}, or Asymmetrical Dampining \cite{spooner2018market}. Since there have been many advancements in the supervised learning domain for LOBs \cite{Zhang_2019}, it would be worthwhile to test those techniques in DRLMM.

	\newpage

	\bibliographystyle{plain}
	\bibliography{bibfile}

	\newpage

	\appendix
	\section{Appendix}
	\subsection{Agent Hyperparameters} \label{AgentConfigurations}
	The hyperparameters used to train agents in the experiment.
	\begin{table}[h	t]
		\centering 
		\begin{tabular}{p{0.02\textwidth} p{0.2\textwidth} p{0.34\textwidth}}
			\toprule\toprule
			\textbf{\#} & \textbf{Parameter} & \textbf{Value} \\ [0.5ex] 
			\hline
			1 & action\_repeat & 5 \\ 
			\hline
			2 & window\_size & 100 \\
			\hline
			3 & transaction\_fees & 0\% for Limit, 0.2\% for Market \\
			\hline
			4 & max\_positions & 10 \\
			\hline
			5 & gamma $\gamma$ & 0.99 \\
			\hline
			6 & learning rate $\alpha$ & 3e-4 \\
			\hline
			7 & LOB levels & 15 \\
			\hline
			8 & n\_steps (PPO) & 256 (PPO) / 40 (A2C) \\
			\hline
			9 & training\_steps & 10,000,000 \\
			\hline
			10 & action\_space & 17 \\
			\hline
		\end{tabular}
		\caption{Hyperparameters for experiments.} 
		\label{table:hyperparameters} 
	\end{table} 
	
	\subsection{Testing Results - Same Currency}
	
	\begin{figure}[h!]
		\includegraphics[scale=0.22]{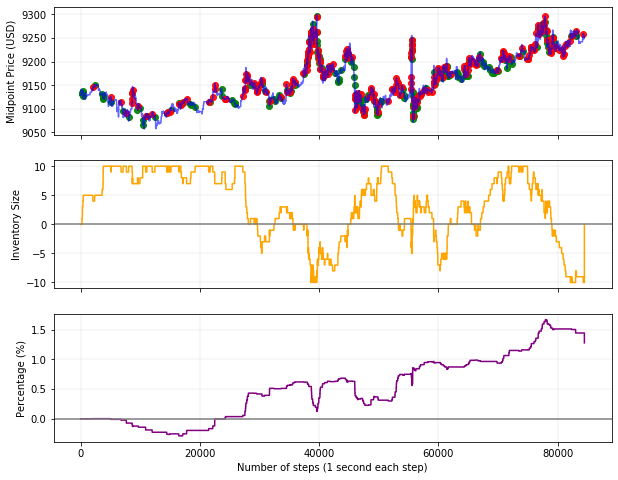}
		\includegraphics[scale=0.22]{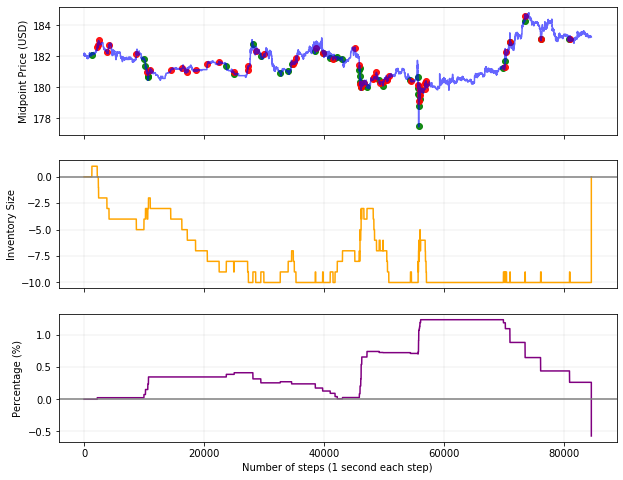}
		\includegraphics[scale=0.22]{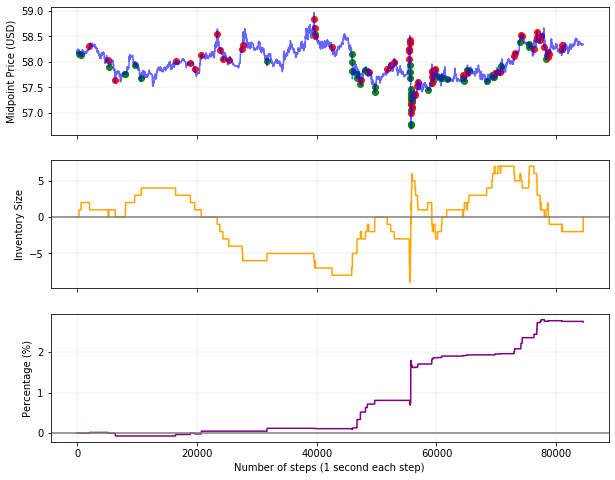}
		\caption{Test results of PPO agents on November 1, 2019. Buy and sell orders executed by the agent are marked in \emph{red} and \emph{green}. \emph{Left:} BTC-USD with action repeat set to 10. \emph{Center:} ETH-USD with action repeat set to 10. \emph{Right:} LTC-USD with action repeat set to 5.}
		\label{fig:ppo_nov_01}
	\end{figure}	
	\begin{figure}[h!]
		\includegraphics[scale=0.22]{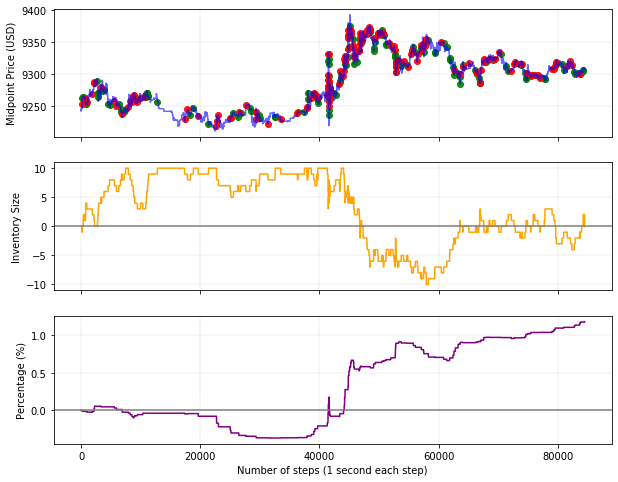}
		\includegraphics[scale=0.22]{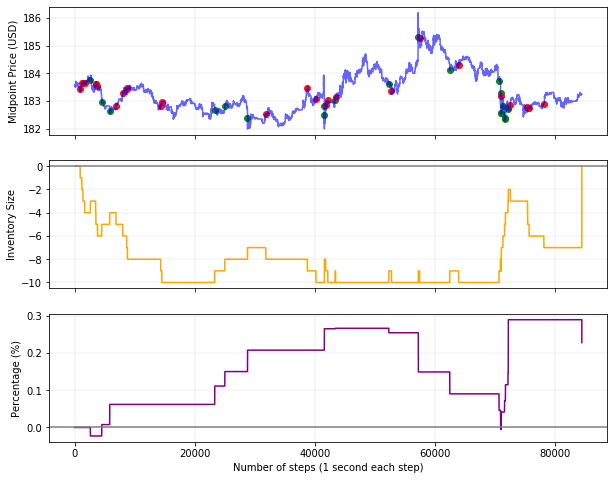}
		\includegraphics[scale=0.22]{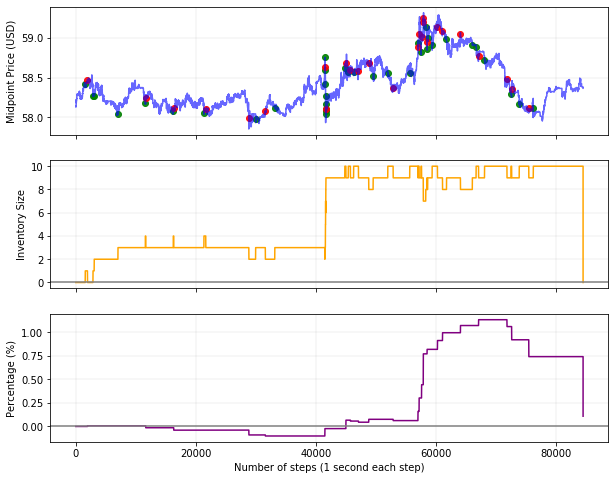}
		\caption{Test results of PPO agents on November 2, 2019. Buy and sell orders executed by the agent are marked in \emph{red} and \emph{green}. \emph{Left:} BTC-USD with action repeat set to 5. \emph{Center:} ETH-USD with action repeat set to 10. \emph{Right:} LTC-USD with action repeat set to 10.}
		\label{fig:ppo_nov_02}
	\end{figure}	
	\begin{figure}[h!]
		\includegraphics[scale=0.22]{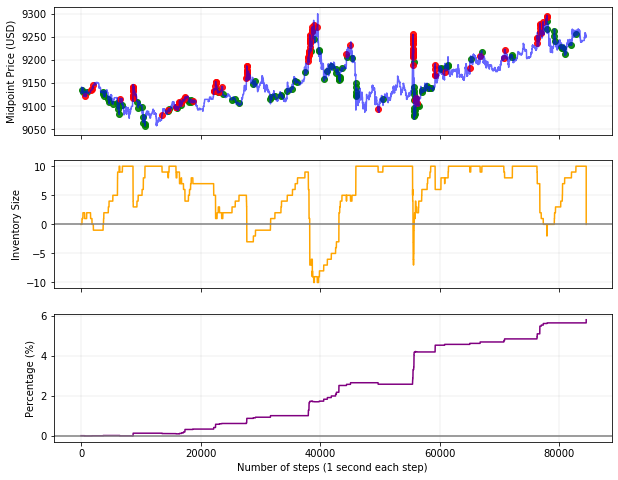}
		\includegraphics[scale=0.22]{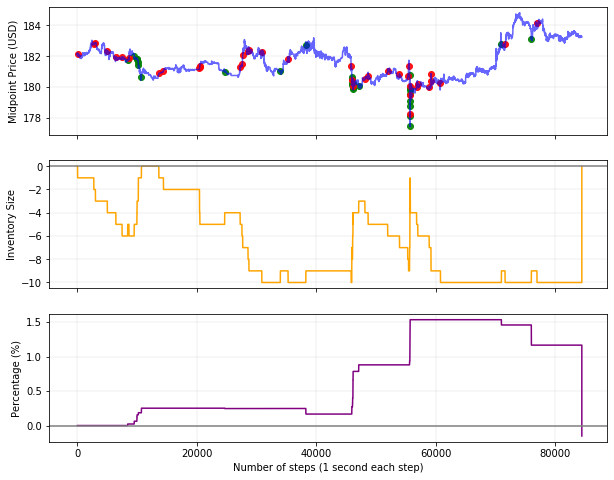}
		\includegraphics[scale=0.22]{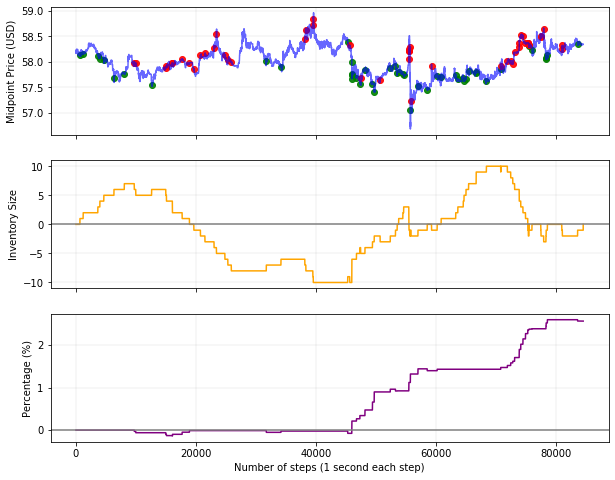}
		\caption{Test results of A2C agents on November 1, 2019. Buy and sell orders executed by the agent are marked in \emph{red} and \emph{green}. \emph{Left:} BTC-USD with action repeat set to 1. \emph{Center:} ETH-USD with action repeat set to 5. \emph{Right:} LTC-USD with action repeat set to 1.}
		\label{fig:a2c_nov_01}
	\end{figure}	
	\begin{figure}[h!]
		\includegraphics[scale=0.22]{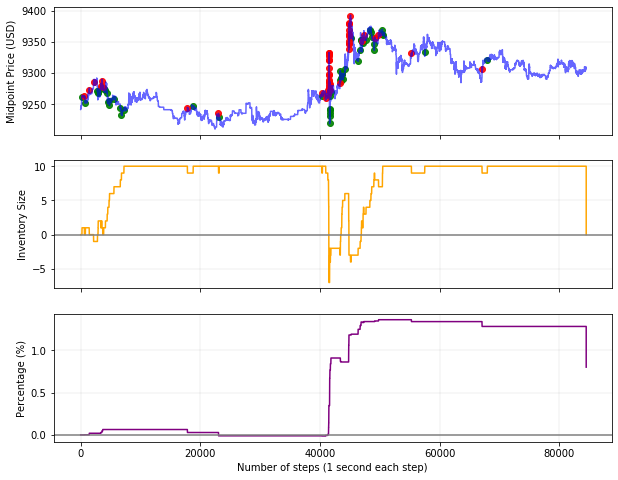}
		\includegraphics[scale=0.22]{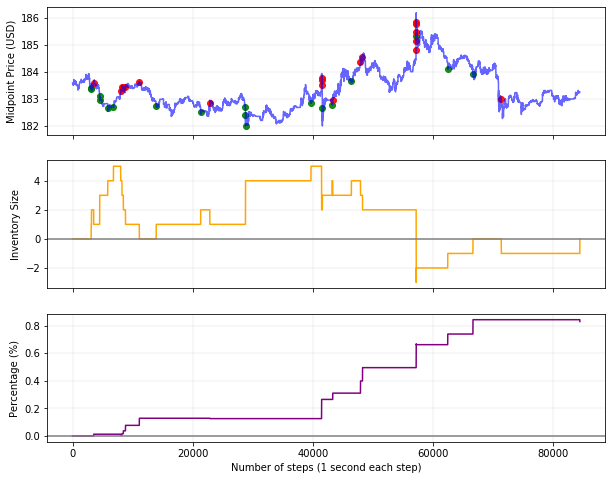}
		\includegraphics[scale=0.22]{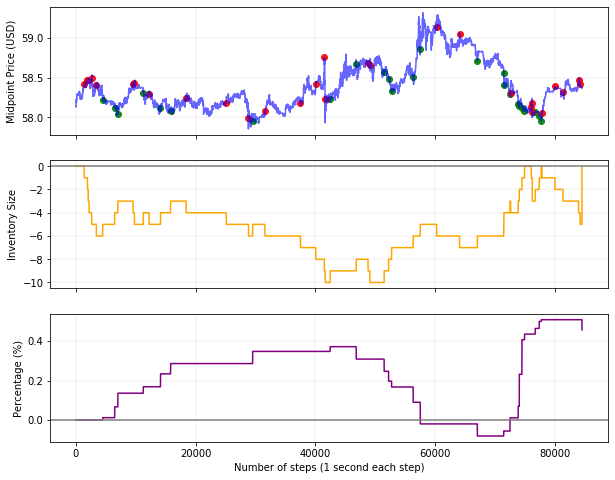}
		\caption{Test results of A2C agents on November 2, 2019. Buy and sell orders executed by the agent are marked in \emph{red} and \emph{green}. \emph{Left:} BTC-USD with action repeat set to 1. \emph{Center:} ETH-USD with action repeat set to 1. \emph{Right:} LTC-USD with action repeat set to 1.}
		\label{fig:a2c_nov_02}
	\end{figure}

	\subsection{Testing Results - Different Currencies}
	
	\begin{figure}[h!]
		\includegraphics[scale=0.22]{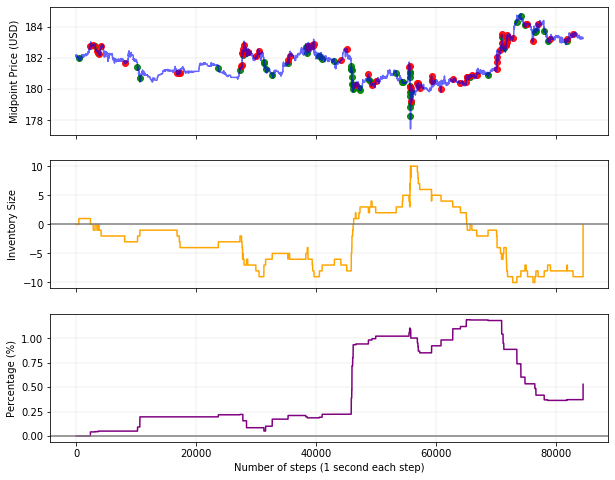}
		\includegraphics[scale=0.22]{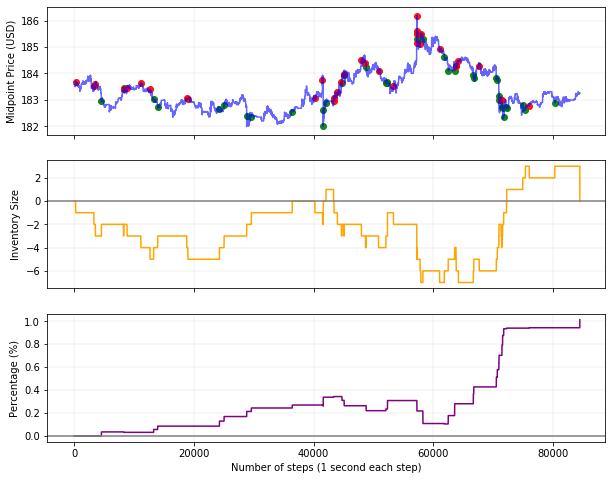}
		\includegraphics[scale=0.22]{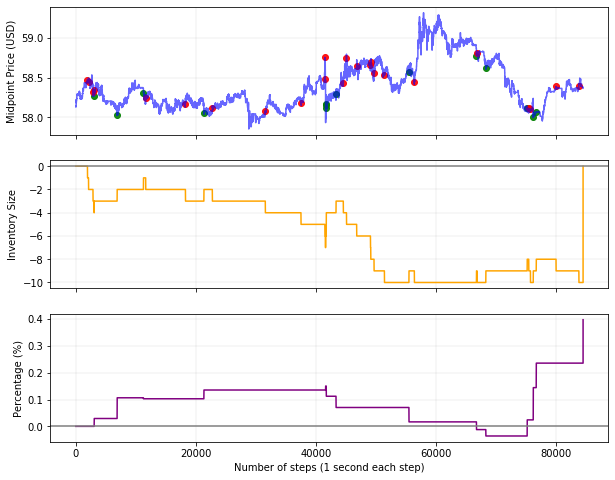}
		\caption{Test results of PPO agents on various test data sets. Buy and sell orders executed by the agent are marked in \emph{red} and \emph{green}. \emph{Left:} BTC-USD trained agent tested on ETH-USD's data from November 1, 2019 with action repeat set to 10. \emph{Center:} BTC-USD trained agent tested on ETH-USD's data from November 2, 2019 with action repeat set to 5. \emph{Right:} ETH-USD trained agent tested on LTC-USD's data from November 2, 2019 with action repeat set to 5.}
		\label{fig:ppo_multi}
	\end{figure}	
	\begin{figure}[h!]
		\includegraphics[scale=0.22]{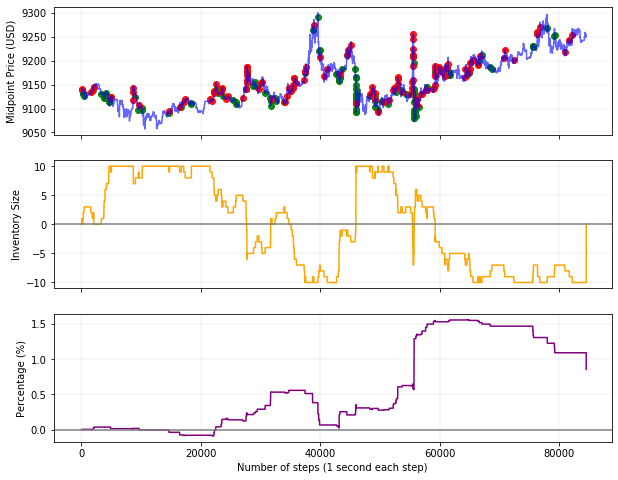}
		\includegraphics[scale=0.22]{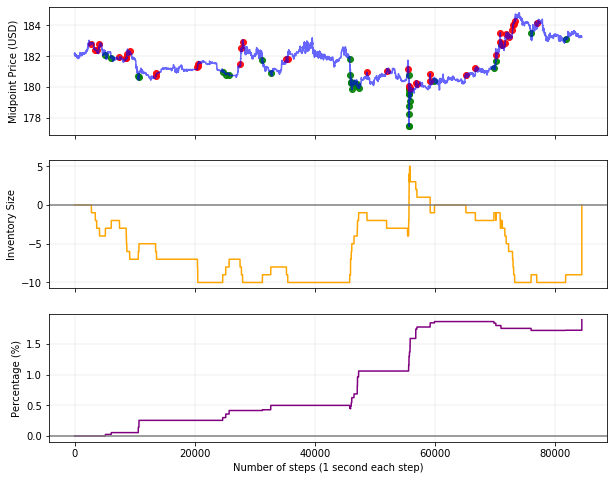}
		\includegraphics[scale=0.22]{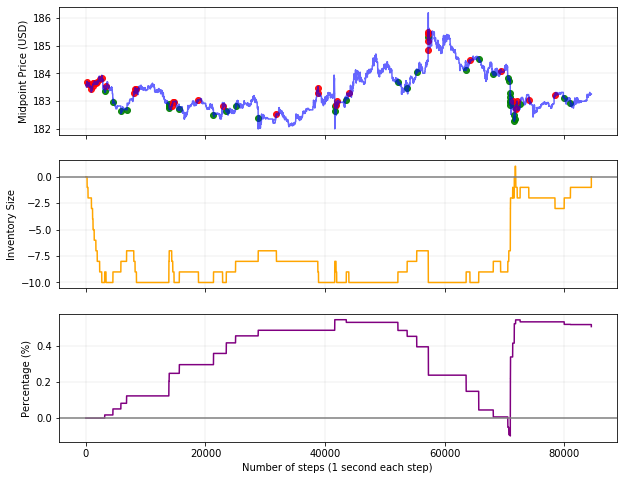}
		\caption{Test results of A2C agents on various test data sets. Buy and sell orders executed by the agent are marked in \emph{red} and \emph{green}. \emph{Left:} ETH-USD trained agent tested on BTC-USD's data from November 1, 2019 with action repeat set to 1. \emph{Center:} BTC-USD trained agent tested on ETH-USD's data from November 1, 2019 with action repeat set to 10. \emph{Right:} LTC-USD trained agent tested on ETH-USD's data from November 2, 2019 with action repeat set to 10.}
		\label{fig:a2c_multi}
	\end{figure}
	
\end{document}